\documentclass[twocolumn,pre]{revtex4}

\usepackage{dcolumn}
\usepackage{amsmath}

\usepackage{graphicx}

\renewcommand{\d}{\mathrm{d}}
\renewcommand{\i}{\mathrm{i}}

\newcommand{\e}{\mathrm{e}}

\newcommand{\av}[1]{\langle#1\rangle}

\newcommand{\defn}{\textit}

\begin{document}

\title{Component sizes in networks with arbitrary degree distributions}
\author{M. E. J. Newman}
\affiliation{Department of Physics,
University of Michigan, Ann Arbor, MI 48109}
\affiliation{Santa Fe Institute, 1399 Hyde Park Road, Santa Fe, NM 87501}
\begin{abstract}
  We give an exact solution for the complete distribution of component
  sizes in random networks with arbitrary degree distributions.  The
  solution tells us the probability that a randomly chosen node belongs to
  a component of size~$s$, for any~$s$.  We apply our results to networks
  with the three most commonly studied degree distributions---Poisson,
  exponential, and power-law---as well as to the calculation of cluster
  sizes for bond percolation on networks, which correspond to the sizes of
  outbreaks of SIR epidemic processes on the same networks.  For the
  particular case of the power-law degree distribution, we show that the
  component size distribution itself follows a power law everywhere below
  the phase transition at which a giant component forms, but takes an
  exponential form when a giant component is present.
\end{abstract}
\maketitle

There has in recent years been considerable interest within the physics
community in the properties of networks~\cite{DM02,Newman03d,NBW06}.
Methods from physics, and particularly from statistical physics, have
proved invaluable for understanding the structure and behavior of networked
systems such as the Internet, the world wide web, metabolic networks,
protein interaction networks, and social networks of interactions between
people.  In particular, by creating simple (and sometimes not-so-simple)
models of network structure and formation, researchers have gained insight
about the way networks behave as a function of the basic parameters
governing their topology.

One of the most fundamental parameters of a network is its degree
distribution.  The degree of a node or vertex in a network is the number of
edges connected to that vertex, and the frequency distribution of the
degrees of vertices has been shown to have a profound influence on almost
every aspect of network structure and function, including path lengths,
clustering, robustness, centrality indices, spreading processes, and many
others.  Various network models have been used to illuminate the effects of
the degree distribution, but perhaps the most widely studied, and certainly
one of the simplest, is the so-called configuration model.

In the configuration model only the degrees of vertices are specified and
nothing else; except for the constraint imposed by the degrees, connections
between vertices are random.  Equivalently, configuration model networks
can be thought of as networks drawn uniformly at random from the set of all
possible networks whose vertices have the specified degrees.  One of the
primary attractions of the configuration model is that many of its
properties can be calculated exactly in the limit of large system size and
for this reason it has become one of the fundamental tools for the
quantitative understanding and study of networks.  In 1995 Molloy and
Reed~\cite{MR95} gave an exact criterion for the existence of a giant
component in the model and later also gave an expression for the expected
size of that component~\cite{MR98}.  Newman, Strogatz, and
Watts~\cite{NSW01} gave additional expressions for a variety of other
properties including number of vertices a given distance from a randomly
chosen vertex, average path length in the giant component, and critical
exponents near the transition at which the giant component appears, as well
as generalizations of the model to bipartite and directed networks, and
many further results have been presented since by a variety of authors.

One fundamental result that has been missing, however, is an expression for
the sizes of components in the model other than the giant component.  More
specifically, if we choose a vertex at random from the network, what is the
probability that it belongs to a component of a given size?  As well as
being a central structural property of the network, this distribution is
directly related to important practical issues such as the distribution of
the sizes of disease outbreaks for diseases spreading over contact
networks~\cite{Grassberger82,Newman02c}.

At first sight, calculation of the component sizes appears difficult.  One
can derive equations that must be satisfied by the generating function for
the distribution of component sizes~\cite{NSW01}, but usually these
equations cannot be solved.  Here we show, however, that it is nonetheless
possible to derive an explicit expression for the complete distribution of
component sizes in the configuration model for general degree distribution.
In particular, we show that it is possible to derive closed-form
expressions for component sizes for the three most commonly studied degree
distributions, the Poisson, exponential, and power-law distributions.  We
also show that the same techniques can be used to calculate the sizes of
percolation clusters for percolation models on networks of arbitrary degree
distribution, a development of some interest because of the close
connection between percolation and epidemic processes.  We explore this
connection in the last part of the paper.

Let $p_k$ be the degree distribution of our network, i.e.,~the probability
that a randomly chosen vertex has degree~$k$.  If rather than a vertex we
choose an edge and follow it to the vertex at one of its ends, then the
number of other edges emerging from that vertex follows a different
distribution, the so-called \defn{excess degree distribution}:
\begin{equation}
q_k = {(k+1) p_{k+1}\over\av{k}},
\label{eq:qk}
\end{equation}
as shown in, for example, Ref.~\cite{NSW01}.  Here $\av{k}=\sum_k kp_k$ is
the average degree in the network.

It will be convenient to introduce the probability generating functions for
the two distributions~$p_k$ and~$q_k$, thus:
\begin{equation}
g_0(z) = \sum_{k=0}^\infty p_k z^k,\quad
g_1(z) = \sum_{k=0}^\infty q_k z^k.
\label{eq:defsg0g1}
\end{equation}
Many of our results are more easily expressed in terms of these generating
functions than directly in terms of the degree distributions.  It will also
be convenient to note that
\begin{equation}
\av{k} = g_0'(1),\qquad g_1(z) = {g_0'(z)\over g_0'(1)},
\label{eq:useful}
\end{equation}
where we have made use of Eq.~\eqref{eq:qk} in the second equality.

Now let us consider the distribution of the sizes of components in our
network.  Every vertex belongs to a component of size at least one (the
vertex itself) and every edge connected to the vertex adds at least one
more vertex to the component, and possibly many, if there are lots of other
vertices that are reachable via that edge.  Let us denote by~$t$ the total
number of vertices reachable via a particular edge, let the probability
distribution of~$t$ be~$\rho_t$, and let the generating function for this
distribution be~$h_1(z)=\sum_t \rho_t z^t$.

The probability that a vertex of degree~$k$ belongs to a component of
size~$s$ is the probability that the numbers of vertices reachable along
each of its~$k$ edges sum to $s-1$.  This probability, which we will denote
$P(s|k)$, is given by
\begin{equation}
P(s|k) = \sum_{t_1=1}^\infty\ldots\sum_{t_k=1}
           \delta\bigl( s-1,{\textstyle\sum_{m=1}^k t_m} \bigr)
           \prod_{m=1}^k \rho_{t_m},
\end{equation}
where $\delta(i,j)$ is the Kronecker delta symbol.  Then the
probability~$\pi_s$ of a randomly chosen vertex belonging to a component of
size~$s$ is $\pi_s = \sum_{k=0}^\infty p_k P(s|k)$ and the corresponding
generating function is
\begin{align}
h_0(z) &= \sum_{s=1}^\infty \pi_s z^s
        = \sum_{s=1}^\infty \sum_{k=0}^\infty p_k P(s|k) z^s \nonumber\\
       &\hspace{-2em} = \sum_{k=0}^\infty p_k \sum_{s=1}^\infty z^s
          \sum_{t_1=1}^\infty\ldots\sum_{t_k=1}
          \delta\bigl( s-1,{\textstyle\sum_{m=1}^k t_m} \bigr)
          \prod_{m=1}^k \rho_{t_m} \nonumber\\
       &\hspace{-2em} = z \sum_{k=0}^\infty p_k
          \sum_{t_1=1}^\infty\ldots\sum_{t_k=1} z^{\sum_m t_m}
          \prod_{m=1}^k \rho_{t_m} \nonumber\\
       &\hspace{-2em} = z \sum_{k=0}^\infty p_k
          \biggl[ \sum_{t=1}^\infty \rho_t z^t \biggr]^k
        = z \sum_{k=0}^\infty p_k [h_1(z)]^k.
\label{eq:h0derive}
\end{align}
But the final sum is simply the generating function~$g_0(z)$,
Eq.~\eqref{eq:defsg0g1}, evaluated at $h_1(z)$, and hence
\begin{equation}
h_0(z) = z g_0(h_1(z)).
\label{eq:h0}
\end{equation}
By a similar argument the generating function~$h_1(z)$ can be shown to
satisfy
\begin{equation}
h_1(z) = z g_1(h_1(z)).
\label{eq:h1}
\end{equation}

Between them, Eqs.~\eqref{eq:h0} and~\eqref{eq:h1} allow us, in principle,
to calculate the entire distribution of cluster sizes in our network given
the degree distribution~$p_k$.  Unfortunately, the self-consistent relation
for~$h_1(z)$, Eq.~\eqref{eq:h1}, is in most cases not solvable and hence we
cannot calculate the value of the generating function.  Surprisingly,
however, we can still calculate the probabilities~$\pi_s$.

Since every component is of size at least~1, the generating function
$h_0(z)$ for the component sizes is of leading order~$z$ (or higher) and
hence contains an overall factor of~$z$.  Dividing out this factor and
differentiating, we can write the probability of belonging to a cluster of
size~$s$ as
\begin{equation}
\pi_s = {1\over (s-1)!} \biggl[ {\d^{s-1}\over\d z^{s-1}}
        \biggl( {h_0(z)\over z} \biggr) \biggr]_{z=0}.
\label{eq:genpi1}
\end{equation}
Using Eq.~\eqref{eq:h0}, this can also be written
\begin{align}
\pi_s &= {1\over (s-1)!} \biggl[ {\d^{s-1}\over\d z^{s-1}}
         g_0(h_1(z)) \biggr]_{z=0} \nonumber\\
      &= {1\over (s-1)!} \biggl[ {\d^{s-2}\over\d z^{s-2}}
         \bigl[ g_0'(h_1(z)) h_1'(z) \bigr] \biggr]_{z=0}.
\label{eq:cmpis1}
\end{align}

This expression can be rewritten using Cauchy's formula for the $n$th
derivative of a function,
\begin{equation}
{\d^n\!f\over\d z^n}\biggr|_{z=z_0} = {n!\over2\pi\i}
  \oint {f(z)\over(z-z_0)^{n+1}} \>\d z,
\label{eq:cauchy2}
\end{equation}
where the integral is around a contour that encloses~$z_0$ in the complex
plane but encloses no poles in~$f(z)$.  Applying this formula to
Eq.~\eqref{eq:cmpis1} with $z_0=0$ we get
\begin{subequations}
\begin{align}
\label{eq:genpi2}
\pi_s &= {1\over2\pi\i(s-1)} \oint {g_0'(h_1(z))\over z^{s-1}} \,
         {\d h_1\over\d z} \,\d z \\
\label{eq:genpi3}
      &= {\av{k}\over2\pi\i(s-1)} \oint {g_1(h_1)\over z^{s-1}}
         \>\d h_1,
\end{align}
\label{eq:genpi}
\end{subequations}
where we have used Eq.~\eqref{eq:useful} to eliminate $g_0'$ in favor
of~$g_1$.  In~\eqref{eq:genpi2} we choose the contour to be an
infinitesimal loop around the origin and, since $h_1(z)$ goes to zero as
$z\to0$, the contour in~\eqref{eq:genpi3} is then also an infinitesimal
loop around the origin.

Now regarding $z$ as a function of~$h_1$, rather than the other way around,
we make use of~\eqref{eq:h1} to eliminate~$z$ and write
\begin{equation}
\pi_s = {\av{k}\over2\pi\i(s-1)} \oint {\bigl[ g_1(h_1) \bigr]^s\over
        h_1^{s-1}} \>\d h_1.
\end{equation}
Applying~\eqref{eq:cauchy2} again we then find that
\begin{equation}
\pi_s = {\av{k}\over(s-1)!} \biggl[ {\d^{s-2}\over\d z^{s-2}}
        \bigl[ g_1(z) \bigr]^s \biggr]_{z=0}.
\label{eq:pi0}
\end{equation}
(An alternative and equivalent way to derive this formula---although a less
transparent one---would be to rearrange Eq.~\eqref{eq:h1} to give~$z$ as a
function of $h_1$ and then apply the Lagrange inversion theorem~\cite{AS65}
to derive the Taylor expansion of $h_1$ or~$h_0$.  Indeed,
Eqs.~\eqref{eq:genpi1} to~\eqref{eq:pi0} are essentially a proof of a
special case of the inversion theorem, as applied to the problem in hand.)

The only exception to Eq.~\eqref{eq:pi0} is for the case $s=1$, for which
Eq.~\eqref{eq:genpi} gives $0/0$ and is therefore clearly incorrect.
However, since the only way to belong to a component of size~1 is to have
no connections to any other vertices, the probability $\pi_1$ is trivially
equal to the probability of having degree zero:
\begin{equation}
\pi_1 = p_0.
\label{eq:pi1}
\end{equation}

Between them, Eqs.~\eqref{eq:pi0} and~\eqref{eq:pi1} give the entire
distribution of component sizes in terms of the degree distribution.  They
tell us explicitly the probability that a randomly chosen vertex belongs to
a component of any given size~$s$.  For any specific choice of degree
distribution, the application of Eq.~\eqref{eq:pi0} still requires us to
perform the derivatives.  Any finite number of derivatives can always be
carried out exactly to give expressions for $\pi_s$ to finite order.  It is
also possible in some cases to find a general formula for any derivative
and so derive a closed-form expression for~$\pi_s$ for general~$s$.  In
particular, it turns out to be possible, as we now show, to find such
closed-form expressions for the three distributions most commonly studied
in the literature, the Poisson, exponential, and power-law distributions.

A network in which edges are placed between vertices uniformly at random
has a Poisson degree distribution
\begin{equation}
p_k = \e^{-c} {c^k\over k!},
\label{eq:poisson}
\end{equation}
where $c$ is the distribution mean.  Such networks have been studied widely
for some decades, most famously by Erd\H{o}s and R\'enyi in the 1950s and
1960s~\cite{ER59,ER60}.  Given Eq.~\eqref{eq:poisson}, it is
straightforward to show that $g_0(z)=g_1(z)=\e^{c(z-1)}$ and the
derivatives in Eq.~\eqref{eq:pi0} can be performed to give
\begin{equation}
\pi_s = {\e^{-cs} (cs)^{s-1}\over s!}.
\label{eq:pipoisson}
\end{equation}
(The same expression also works for the special case $s=1$.)  This
expression for the component size distribution of the Poisson random graph
has been derived in the past by a number of other methods---see for
instance~\cite{Bollobas01}---but it is a useful check on our methods to see
it appear here as a special case of the more general formulation.

Few real-world networks, however, have Poisson degree distributions.  Most
have highly right-skewed distributions in which most vertices have low
degree and a small number of ``hubs'' have higher degree.  A number of
networks, for example, are observed to have exponential degree
distributions or distributions with an exponential tail.  Examples include
food webs, power grids, and some social networks~\cite{ASBS00,DWM02a}.
Consider the exponential distribution $p_k = C \e^{-\lambda k}$, where $C$
is the appropriate normalizing constant.  The generating functions in this
case are
\begin{equation}
g_0(z) = {\e^\lambda - 1\over\e^\lambda - z},\qquad
g_1(z) = \biggl[ {\e^\lambda - 1\over\e^\lambda - z} \biggr]^2.
\end{equation}
Again the derivatives are straightforward to carry out and we find that
\begin{equation}
{\d^n\over\d z^n} \bigl[ g_1(z) \bigr]^s
  = {(2s-1+n)!\over(2s-1)!}\,
    {\bigl[ g_1(z) \bigr]^s\over(\e^\lambda-z)^n},
\end{equation}
and hence
\begin{equation}
\pi_s = {(3s-3)!\over(s-1)!(2s-1)!} \, \e^{-\lambda(s-1)}
        \bigl(1-\e^{-\lambda}\bigr)^{2s-1}.
\label{eq:piexp}
\end{equation}
Applying Stirling's approximation for large~$s$ we can show that this
distribution behaves asymptotically as $\pi_s \sim s \e^{-\mu s}$, where
$\mu=2\ln\bigl[\frac32 (1-\e^{-\lambda})\bigr]-\lambda$.  Thus the
component size distribution approximately follows an exponential law
itself, although with an extra leading factor of~$s$ and a different
exponential constant.

However, perhaps the greatest amount of attention in recent years has been
focused on networks that have power-law degree distributions of the form
$p_k\propto k^{-\alpha}$ for some constant
exponent~$\alpha$~\cite{AJB99,FFF99,Kleinberg99b}.  A number of networks
appear to follow this pattern, at least approximately, including the
world wide web, the Internet, citation networks, and some social and
biological networks~\cite{DM02}.  The observed value of the exponent
typically lies in the range $2<\alpha<3$.  Equivalently, we could say that
the excess degree distribution~$q_k$---which appears in the fundamental
formula~\eqref{eq:pi0} via its generating function---follows a power law
with exponent $\alpha-1$.

In fact, in essentially all cases, the observed power law holds only in the
tail of the distribution; the distribution follows some other law for small
degrees.  This leaves us considerable latitude about the distribution we
use in our calculations.  Here we use a so-called Yule distribution
for~$q_k$, with a typical real-world value of $\alpha=2.5$ for the
exponent:
\begin{equation}
q_k = C {\Gamma(k+\frac12)\over\Gamma(k+2)},
\end{equation}
where $\Gamma(x)$ is the standard gamma function and $C$ is again a
normalizing constant.  It is straightforward to show (by Stirling's
approximation) that this distribution asymptotically follows a power law
$q_k\sim k^{-3/2}$, which corresponds to a raw degree distribution $p_k\sim
k^{-5/2}$.  The Yule distribution appears in a number of contexts in the
study of networks, particularly in the solutions of preferential attachment
models that may explain the origin of power laws in some
networks~\cite{DMS00,KRL00}, and is considered by some to be the most
natural choice of power-law form for discrete distributions.  Employing
this particular choice for our configuration model gives
\begin{equation}
g_1(z) = {1\over 1+\sqrt{1-z}},
\end{equation}
which in turn gives
\begin{align}
\biggl[ {\d^n\over\d z^n} \bigl[ g_1(z) \bigr]^s \biggr]_{z=0}
  &= {2^{-(2n+s)}\over(s-1)!} \times \nonumber\\
  &\hspace{-2em} \sum_{j=0}^{n-1} {(n-1+j)! (s+n-1-j)!\over j!(n-1-j)!}.
\end{align}
Setting $n=s-2$ and substituting into Eq.~\eqref{eq:pi0}, we can complete
the remaining sum to get
\begin{equation}
\pi_s = [1-\ln 2]^{-1} {(3s-5)!\over(s-1)!(2s-2)!}\,s\,2^{3-3s}.
\label{eq:pipl}
\end{equation}

\begin{figure}
\begin{center}
\includegraphics[width=8.3cm]{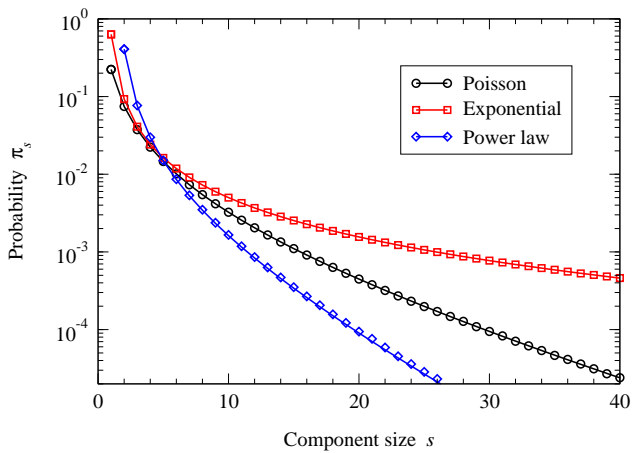}
\end{center}
\caption{The distribution of component sizes in random graphs with Poisson
  ($c=1.5$), exponential ($\lambda=1$), and power-law ($\alpha=2.5$) degree
  distributions.  Solid lines indicate the exact solutions derived in this
  paper.  Points are the results of computer simulations for the same
  degree distributions.  Each point is an average over $5000$ networks of
  $10^6$ vertices each.  Error bars have been omitted, but are smaller than
  the data points in each case.}
\label{fig:results}
\end{figure}

In Fig.~\ref{fig:results} we show the form of this distribution, along with
those for the Poisson and exponential networks, Eqs.~\eqref{eq:pipoisson}
and~\eqref{eq:piexp}.  Also shown in the figure are numerical results for
the distributions of component sizes measured on computer generated
networks with the same degree distributions.  As the figure shows, there is
excellent agreement between the simulations and the exact calculations.

As with the exponential network, we can study the asymptotic form of the
component size distribution~\eqref{eq:pipl} for the power-law network by
making use of Stirling's approximation.  We find that in the limit of
large~$s$, $\pi_s \sim s^3 \e^{-\nu s}$, where
$\nu=5\ln2-3\ln3\simeq0.1699\ldots$\ \ Thus again we have an exponential
tail to the distribution.

This last result is at first slightly surprising.  One might imagine that
the component size distribution should itself fall off as a power law or
slower because the degree of a vertex provides a lower bound on the size of
the component to which the vertex belongs---the fraction of vertices in
components of size~$s$ or greater must be at least as large as the fraction
of vertices of degree~$s$ or greater and hence the cumulative distribution
of components falls off as slow or slower than the cumulative distribution
of degrees.

So how is it possible that we have an exponential distribution of component
sizes in the present case?  The answer is that we are studying a network
that has a giant component.  Vertices not in the giant component---which
make up almost all of the component size distribution---have a different
degree distribution from the graph as a whole because the probability of
not being in the giant component dwindles exponentially with increasing
degree~\cite{Newman02c}.  This creates an exponential cutoff for the degree
distribution, and hence we are back to the situation we had for the
exponential network, which gave an exponential component size distribution.

Thus in a power-law network we expect $\pi_s$ to have an exponential tail
whenever there is a giant component in the network, but a power-law tail
when there is no giant component.  This contrasts with the case for
essentially every other degree distribution, where we expect a power-law
distribution of component sizes only precisely at the phase transition
where the giant component forms; everywhere else we expect the distribution
to fall off exponentially or faster~\cite{NSW01}.

The methods described here can be extended to the calculation of cluster
sizes for percolation processes on networks also.  Of particular interest
is the bond percolation process, whose cluster sizes give the distribution
of outbreaks for a standard SIR epidemiological process on the same
network~\cite{Mollison77,Grassberger82}.  Bond percolation can be framed in
the same language as the calculation of component sizes above by
considering the network formed by just the occupied edges.  If the
occupation probability is~$\phi$, then it is straightforward to
show~\cite{Newman02c} that the generating functions for the degree
distribution and excess degree distribution of this latter network are
$g_0(1-\phi+\phi z)$ and $g_1(1-\phi+\phi z)$, with $g_0$ and $g_1$ defined
as before.  Substituting into Eq.~\eqref{eq:pi0}, we then find
\begin{equation}
\pi_s = {\phi^{s-1}\av{k}\over(s-1)!} \biggl[ {\d^{s-2}\over\d z^{s-2}}
        \bigl[ g_1(z) \bigr]^s \biggr]_{z=1-\phi}.
\end{equation}

This result immediately implies that for all $\phi<1$ the distribution of
cluster sizes falls off at least exponentially with increasing~$s$.  Thus,
in the language of epidemiology, we will never see a power-law distribution
of outbreak sizes, even if the network has a power-law degree distribution.
This is, overall, good news: it implies that there will be no fat tail to
the outbreak distribution and hence no unexpectedly large outbreaks,
regardless of whether the network has a giant component.

To conclude, we have given an exact solution for the distribution of
component sizes in random graphs with arbitrary degree distributions and
applied it to networks with Poisson, exponential, and power-law distributed
degrees.  In the latter case we find that though the network has a
power-law distribution of component sizes when there is no giant component,
the distribution develops an exponential tail once a giant component
appears.  We have also applied our methods to bond percolation on networks,
finding that percolation clusters always have an exponential tail to their
distribution whenever the bond occupation probability is less than one.

The author thanks Cris Moore for useful conversations.  This work was
funded in part by the National Science Foundation under grant DMS--0405348
and by the Santa Fe Institute.


\begin{thebibliography}{10}
\expandafter\ifx\csname url\endcsname\relax
  \def\url#1{\texttt{#1}}\fi
\expandafter\ifx\csname urlprefix\endcsname\relax\def\urlprefix{URL }\fi

\bibitem{DM02}
S.~N. Dorogovtsev and J.~F.~F. Mendes, Evolution of networks. \textit{Advances
  in Physics} \textbf{51}, 1079--1187 (2002).

\bibitem{Newman03d}
M.~E.~J. Newman, The structure and function of complex networks. \textit{SIAM
  Review} \textbf{45}, 167--256 (2003).

\bibitem{NBW06}
M.~E.~J. Newman, A.-L. Barab\'asi, and D.~J. Watts, \textit{The Structure and
  Dynamics of Networks}. Princeton University Press, Princeton (2006).

\bibitem{MR95}
M.~Molloy and B.~Reed, A critical point for random graphs with a given degree
  sequence. \textit{Random Structures and Algorithms} \textbf{6}, 161--179
  (1995).

\bibitem{MR98}
M.~Molloy and B.~Reed, The size of the giant component of a random graph with a
  given degree sequence. \textit{Combinatorics, Probability and Computing}
  \textbf{7}, 295--305 (1998).

\bibitem{NSW01}
M.~E.~J. Newman, S.~H. Strogatz, and D.~J. Watts, Random graphs with arbitrary
  degree distributions and their applications. \textit{Phys. Rev. E}
  \textbf{64}, 026118 (2001).

\bibitem{Grassberger82}
P.~Grassberger, On the critical behavior of the general epidemic process and
  dynamical percolation. \textit{Math. Biosci.} \textbf{63}, 157--172 (1982).

\bibitem{Newman02c}
M.~E.~J. Newman, Spread of epidemic disease on networks. \textit{Phys. Rev. E}
  \textbf{66}, 016128 (2002).

\bibitem{AS65}
M.~Abramowitz and I.~A. Stegun (eds.), \textit{Handbook of Mathematical
  Functions}. Dover Publishing, New York (1974).

\bibitem{ER59}
P.~Erd\H{o}s and A.~R\'enyi, On random graphs. \textit{Publicationes
  Mathematicae} \textbf{6}, 290--297 (1959).

\bibitem{ER60}
P.~Erd\H{o}s and A.~R\'enyi, On the evolution of random graphs.
  \textit{Publications of the Mathematical Institute of the Hungarian Academy
  of Sciences} \textbf{5}, 17--61 (1960).

\bibitem{Bollobas01}
B.~Bollob\'as, \textit{Random Graphs}. Academic Press, New York, 2nd edition
  (2001).

\bibitem{ASBS00}
L.~A.~N. Amaral, A.~Scala, M.~Barth\'el\'emy, and H.~E. Stanley, Classes of
  small-world networks. \textit{Proc. Natl. Acad. Sci. USA} \textbf{97},
  11149--11152 (2000).

\bibitem{DWM02a}
J.~A. Dunne, R.~J. Williams, and N.~D. Martinez, Food-web structure and network
  theory: The role of connectance and size. \textit{Proc. Natl. Acad. Sci. USA}
  \textbf{99}, 12917--12922 (2002).

\bibitem{AJB99}
R.~Albert, H.~Jeong, and A.-L. Barab\'asi, Diameter of the world-wide web.
  \textit{Nature} \textbf{401}, 130--131 (1999).

\bibitem{FFF99}
M.~Faloutsos, P.~Faloutsos, and C.~Faloutsos, On power-law relationships of the
  internet topology. \textit{Computer Communications Review} \textbf{29},
  251--262 (1999).

\bibitem{Kleinberg99b}
J.~M. Kleinberg, S.~R. Kumar, P.~Raghavan, S.~Rajagopalan, and A.~Tomkins, The
  {W}eb as a graph: Measurements, models and methods. In T.~Asano, H.~Imai,
  D.~T. Lee, S.-I. Nakano, and T.~Tokuyama (eds.), \textit{Proceedings of the
  5th Annual International Conference on Combinatorics and Computing}, number
  1627 in Lecture Notes in Computer Science, pp. 1--18, Springer, Berlin
  (1999).

\bibitem{DMS00}
S.~N. Dorogovtsev, J.~F.~F. Mendes, and A.~N. Samukhin, Structure of growing
  networks with preferential linking. \textit{Phys. Rev. Lett.} \textbf{85},
  4633--4636 (2000).

\bibitem{KRL00}
P.~L. Krapivsky, S.~Redner, and F.~Leyvraz, Connectivity of growing random
  networks. \textit{Phys. Rev. Lett.} \textbf{85}, 4629--4632 (2000).

\bibitem{Mollison77}
D.~Mollison, Spatial contact models for ecological and epidemic spread.
  \textit{Journal of the Royal Statistical Society B} \textbf{39}, 283--326
  (1977).

\end{thebibliography}
\end{document}